\documentclass[aps,pre,twocolumn,showpacs]{revtex4}
\usepackage{epsfig}
\usepackage{times}
\begin{document}

\title{Topology independent impact of noise on cooperation in spatial public goods games}

\author{Attila Szolnoki$^1$, Matja{\v z} Perc$^2$, and Gy{\"o}rgy Szab{\'o}$^1$}
\affiliation
{$^1$Research Institute for Technical Physics and Materials Science,
P.O. Box 49, H-1525 Budapest, Hungary \\
$^2$Department of Physics, Faculty of Natural Sciences and Mathematics, University of \\ Maribor, Koro{\v s}ka cesta 160, SI-2000 Maribor, Slovenia}

\begin{abstract}
We study the evolution of cooperation in public goods games on different regular graphs as a function of the noise level underlying strategy adoptions. We focus on the effects that are brought about by different group sizes of public goods games in which individuals participate, revealing that larger groups of players may induce qualitatively different behavior when approaching the deterministic limit of strategy adoption. While by pairwise interactions an intermediate uncertainty by strategy adoptions may ensure optimal conditions for the survival of cooperators at a specific graph topology, larger groups warrant this only in the vicinity of the deterministic limit independently from the underlying graph. These discrepancies are attributed to the indirect linkage of otherwise not directly connected players, which is brought about by joint memberships within the larger groups. Thus, we show that increasing the group size may introduce an effective transition of the interaction topology, and that the latter shapes the noise dependence of the evolution of cooperation in case of pairwise interactions only.
\end{abstract}

\maketitle

\section{Introduction}

Public goods are prone to exploitations by individuals who want to reap benefits on the expense of others. Such situations, in which actions ensuring or enhancing individual prosperity harm collective wellbeing, are known as social dilemmas \cite{macy_pnas02}. Cooperation is regarded as the strategy leading away from the impending social decline \cite{axelrod_84}. However, the evolution of cooperation is at odds with the defecting strategy, which promises higher individual payoffs and is thus more likely to spread in a success-driven environment where the evolutionary process is governed by the imitation of the more successful strategy. Indeed, sustenance of cooperation within groups of selfish individuals is a challenge faced by scientists across different fields of research ranging from sociology and ecology to economics \cite{nowak_s06}. The subtleties of cooperation within groups of selfish individuals are most frequently investigated within the framework of evolutionary game theory \cite{hofbauer_88, gintis_00, nowak_06, szabo_pr07}. While the prisoner's dilemma game is unrivaled in popularity when it comes to studying the evolution of cooperation through pairwise interactions, the closely related public goods game accounts for group interactions as well. Within the latter cooperators contribute to the public good whereas defectors do not. All contributions are summed and multiplied by an enhancement factor $r>1$, and subsequently equally divided among all players irrespective of their strategies. Thus, defectors bear no costs when collecting identical benefits as cooperators, which ultimately results in widespread defection. This theoretical prediction, however, disagrees with experimental findings \cite{fehr_ars07}, and accordingly, several mechanisms have been proposed that facilitate the evolution of cooperation in public goods games. Punishment in particular seems to be a viable route to cooperative behavior \cite{fehr_n02, boyd_pnas03, brandt_pnas06}, but it raises the so-called second-order free rider problem where cooperators exploit the additional investments of punishers \cite{second_order}. Furthermore, the effectiveness of such actions depends on whether the participation in public goods game is optional or not \cite{hauert_s07}. Social diversity \cite{santos_n08} and volunteering \cite{hauert_s02} may also promote cooperation in public goods games, as does the relaxation of strategy adoption criteria to account for mutation and random exploration of available strategies \cite{traulsen_pnas09}.

Apart from the mainstream efforts focusing on the identification of mechanisms promoting cooperation, the public goods game attracted considerable interest also from the viewpoint of phenomena that are of interest to physicists, especially in view of spatial public goods games \cite{brandt_prsb03}, being interesting and realistic extensions of the well-mixed case. Phase transitions in spatial public goods games have been investigated in \cite{szabo_prl02}. It was shown that pattern formation, previously identified in the contexts of excitable media for example \cite{sagues_rmp07}, may also underly the evolution of ecological public goods \cite{wakano_pnas09}. Public goods games have also been studied in different geometries in view of the importance of voluntary and compulsory interactions \cite{hauert_c03}, whereas the impact of continuous spatial structure on the success of altruistic behavior has been investigated in \cite{wakano_mbs06}. Effects of inhomogeneous player activities and noise on cooperation in spatial public goods games have been examined in \cite{guan_pre07}. Importantly, it has been reported that noise, originating from mistakes in strategy adoptions or misinterpretations of individual fitness \cite{szabo_pre98}, plays a crucial role by the evolution of cooperation, resulting even in a non-monotonous outlay of cooperator density in dependence on the noise intensity. Qualitatively similar findings were reported for the impact of diversity on cooperation in spatial public goods games \cite{yang_pre09}. This is in agreement with results reported earlier for the spatial prisoner's dilemma game \cite{vukov_pre06, perc_njp06}, although there the particular structure of the interaction topology has been found crucial for the noise dependence of the evolution of cooperation.

\begin{figure}
\centerline{\epsfig{file=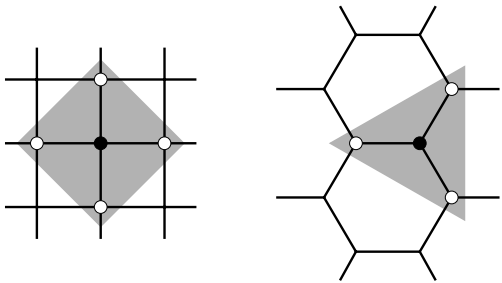,width=7cm}}
\caption{Schematic presentation of employed spatial lattices that lack overlapping triangles. Left: square lattice. Right: honeycomb lattice. In both panels the gray shaded region encompasses players that engage in a round of the public goods game. Due to the extension of the group size $G$ from two (pairwise interaction; $G=2$) to five (left) or four (right) players, individuals denoted by white circles surrounding the focal player (black circle) become effectively connected via the public goods game.}
\label{scheme}
\end{figure}

Here we aim to examine whether the interaction topology by public goods games plays a similarly crucial role. We therefore consider public goods games on different types of regular graphs, whereby focusing specifically on the impact of different group sizes of players participating in every instance of the game. Notably, we focus on regular graphs (where each node is connected to $z$ neighbors) to avoid effects originating from heterogeneous interaction structures \cite{santos_n08}. First, we employ square and honeycomb spatial lattices representing regular graphs where the overlapping triangles are missing. Accordingly, these interaction topologies have a clustering coefficient equal to zero. The latter structural feature has been found crucial for the evolution of cooperation within the prisoner's dilemma game when approaching the deterministic limit of strategy adoption \cite{szabo_pre05, vukov_pre06, szabo_pr07}. In particular, while in the absence of overlapping triangles cooperators fail to survive in the deterministic limit, cooperative behavior thrives in the opposite case, \textit{i.e.} when the interaction structure contains overlapping triangles, as is the case by triangular ($z=6$) or kagome ($z=4$) lattices that we consider also presently. The premise of this work relies on the assumption that abandoning pairwise interactions, as is commonly the case by the prisoner's dilemma game, and allowing for the more general group interactions within the framework of public goods games can lead to qualitatively different results. Figure~\ref{scheme} features a schematic representation of the square and honeycomb lattice, whereby the gray shaded areas encompass $G=z+1$ players belonging to the same group when engaging in a round of the public goods game. Evidently, due to the joint membership there appear indirect links between otherwise not directly connected players, which in turn introduce effective overlapping triangles to the interaction structure. In what follows, we will investigate the impact of this fact on the noise dependence of the evolution of cooperation in spatial public goods games.

The remainder of this paper is organized as follows. In the next section we describe the employed evolutionary public goods game, and in Section III we present the results. Lastly, we summarize the main findings and compare them with results reported earlier for other spatial evolutionary games.

\section{Public goods game}

The public goods game is staged on a regular interaction graph, as schematically depicted in Fig.~\ref{scheme}, whereon initially each player on site $x$ is designated either as a cooperator ($s_x = C$) or as a defector ($s_x = D$) with equal probability. Depending on the group size $G$, each selected player $x$ acquires its payoff $P_x$ by accumulating its share of the public good from all groups with which it is affiliated (\textit{e.g.} if $G=5$ there exist five such groups on the square lattice; see Fig.~\ref{scheme}), whereby cooperators contribute to the public good with an asset $a=1$ while defectors contribute nothing. Subsequently, the total contribution that accumulates within a group is multiplied by the enhancement factor $r$ and divided equally among all players irrespective of their strategy. Employing the Monte Carlo simulation procedure, each elementary step involves randomly selecting one player $x$ and one of its neighbors $y$. Following the accumulation of payoffs $P_x$ and $P_y$ as described above, player $x$ tries to enforce its strategy $s_x$ on player $y$ in accordance with the probability
\begin{equation}
W(s_x \rightarrow s_y)=\frac{1}{1+\exp[(P_y-P_x)/K]},
\label{fermi}
\end{equation}
where $K$ denotes the uncertainty by strategy adoptions \cite{szabo_pr07}. In $K \to 0$ limit player $x$ always succeeds in enforcing its strategy to player $y$ if only $P_x > P_y$ but never otherwise. For $K > 0$, however, strategies performing worse may also be adopted based on unpredictable variations in payoffs or errors in the decision making, for example. During a Monte Carlo step (MCS) all players will have a chance to pass their strategy once on average. For the results presented below we used lattices having $L=400$ to $1600$ linear size and up to $10^5$ MCS before determining the fraction of cooperators $\rho_C$ within the whole population. Importantly, to account for different group sizes affecting the absolute values of the payoffs, and thus indirectly influencing also $r$ and $K$, the latter two parameters must be considered normalized with $G$ to ensure relevant comparisons of results, similarly as has been done recently in \cite{santos_n08}.

\section{Results}

\begin{figure}
\centerline{\epsfig{file=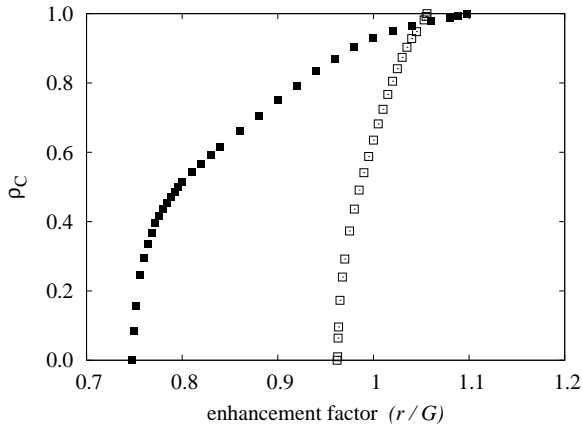,width=8.5cm}}
\caption{Density of cooperators $\rho_C$ in dependence on the normalized (see main text for details) enhancement factor $r$ for the square lattice at $G=2$ (open squares) and $G=5$ (filed squares). Note that large groups can sustain cooperation by lower enhancement factors than small groups. The uncertainty by strategy adoptions was $K/G=0.1$.}
\label{cross}
\end{figure}

We start by presenting results obtained on the square lattice for $G=2$ (pairwise interactions) and $G=5$ in Fig.~\ref{cross}. The dependence of $\rho_C$ on the normalized enhancement factor $r$ reveals that larger groups are able to sustain cooperative behavior by lower values of $r/G$ than smaller groups. In agreement with the fact that cooperation is more difficult to sustain for lower enhancement factors, we may argue that larger groups facilitate the survival of cooperators. On the other hand, $r/G$ required for complete cooperator dominance rises slightly as the group size is enlarged, which in turn relativises the preceding statement. Irrespective of this, results presented in Fig.~\ref{cross} evidence clearly that group size plays a decisive role by the evolution of cooperation in the public goods game on  the square lattice. Notably, we obtained qualitatively identical results for the honeycomb lattice (not shown).

\begin{figure}
\centerline{\epsfig{file=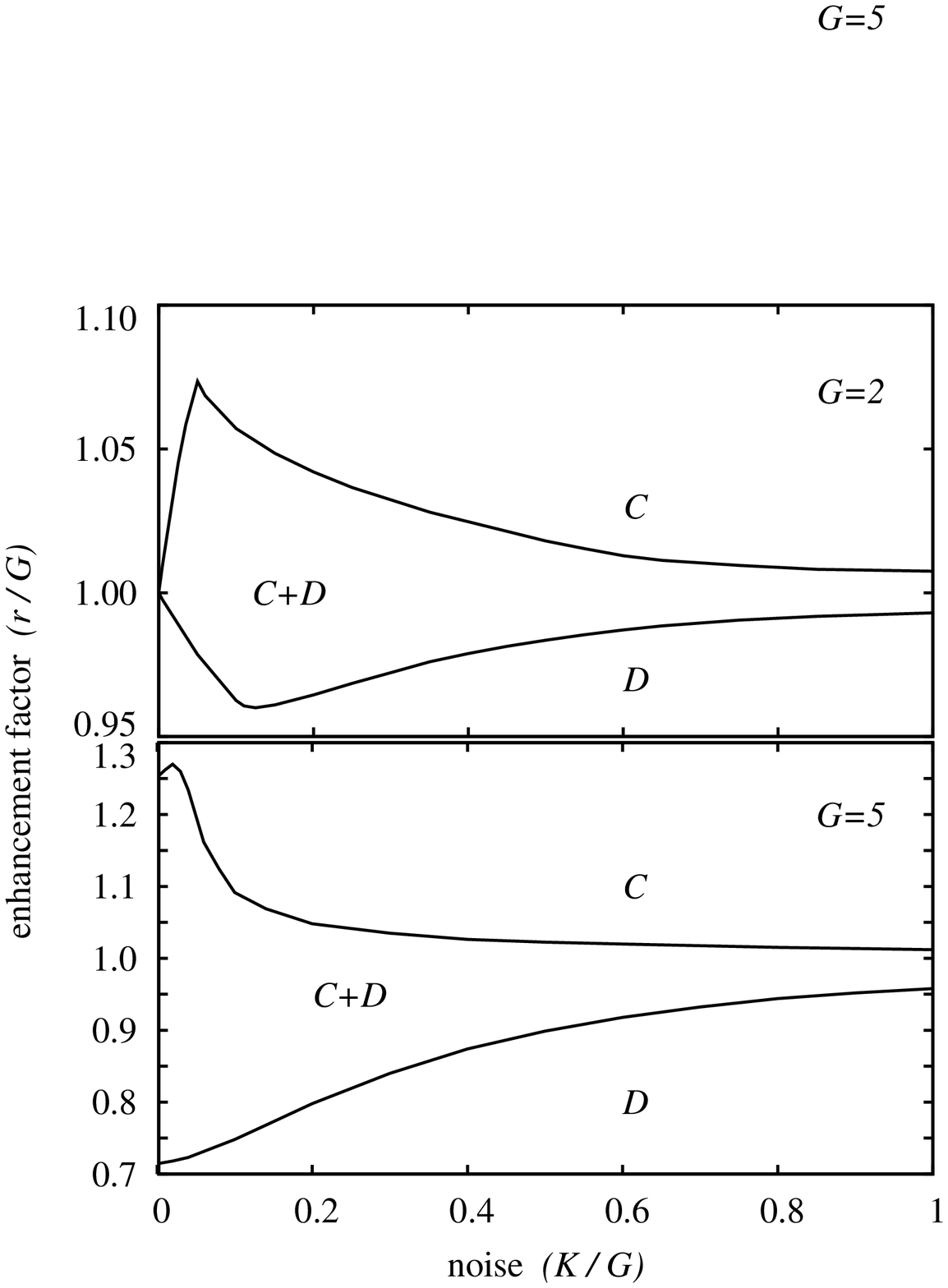,width=8.5cm}}
\caption{$K-r$ phase diagrams for the square lattice at $G=2$ (top) and $G=5$ (bottom). The upper (lower) line denotes the phase boundary between the mixed $C+D$ and homogeneous $C$ ($D$) phases. Note the qualitative difference of the lower phase boundary as it approaches the $K \to 0$ limit, emerging due to the application of different group sizes $G$.}
\label{square}
\end{figure}

Elaborating further on the impact of different group sizes on the evolution of cooperation, we present in Fig.~\ref{square} the full $K-r$ phase diagram for the square lattice at $G=2$ (top) and $G=5$ (bottom). As could be concluded already from results presented in Fig.~\ref{cross}, larger groups indeed lower the enhancement factor that is needed to sustain cooperation. The phase diagram reveals further that this is true irrespective of $K$. Moreover, it is also evident that $r/G$ required for $\rho_C=1$ rises with larger $G$, and that this trend is also largely independent of $K$. The most interesting feature, however, is the qualitatively different behavior of the phase boundaries when approaching the $K \to 0$ limit. Focusing on the lower phase boundary separating the mixed $C+D$ and pure $D$ phases, it is easy to observe that at $G=2$ there exist an intermediate value of $K$ by which the lowest $r$ still ensures the survival of at least some cooperators. The final result in the strong ($K \to 0$) and weak ($K \to \infty$) selection limit is identical, yielding $r=G$ as the phase transition point between $C$ and $D$ phases. Although brought about by a different spatiotemporal evolution of the two strategies, the application of the two extreme noise levels wipes out the impact of spatiality if pairwise interactions ($G=2$) are considered. On the other hand, at $G=5$ the trend of the lower curve is monotonically descending towards the $K \to 0$ limit, which is in sharp contrast with the $G=2$ case. Indeed, while the latter implies the existence of an optimal level of uncertainty for the evolution of cooperation, as was previously reported in \cite{vukov_pre06, perc_njp06} for the prisoner's dilemma game, the $G=5$ case lacks this feature, in turn posing questions with respect to the role of the interaction structure. In particular, the inverted bell-shaped outlay of the critical $r/G$ needed to sustain cooperation in the top panel of Fig.~\ref{cross} can be seen equivalent to the bell-shaped outlay of the critical temptation to defect in dependence on $K$ reported earlier for the prisoner's dilemma game (see \textit{e.g.} Fig. 32 in \cite{szabo_pr07}). However, for the prisoner's dilemma game this feature can be observed only if the interaction graph lacks overlapping triangles, whereas for lattices like the kagome, or the regular graph consisting of overlapping triangles, the bell-shaped dependence vanishes and becomes qualitatively identical to the monotonous outlay of the critical $r/G$ depicted in the bottom panel of Fig.~\ref{square}. Thus, the conclusion imposes that the increase in the group size from $G=2$ to $G=5$ on the square lattice effectively alters the interaction topology in that the joint membership in the larger groups indirectly links the otherwise not linked players.

In order to strengthen the above-outlined reasoning, we have studied the full $K-r$ phase diagram for the honeycomb lattice at $G=2$ and $G=4$ as well. As expected, the lack of overlapping triangles results in an inverted bell-shaped dependence of the critical $r/G$ values constituting the $C+D \rightarrow D$ border line in case of pairwise interactions ($G=2$). More importantly still, the increase of the group size from $G=2$ to $G=4$ induces an identical qualitative change in the outlay of the critical $r/G$ values when approaching the $K \to 0$ limit as we have observed for the square lattice in Fig.~\ref{square}. In particular, the phase boundary between the mixed $C+D$ and homogeneous $D$ phases is monotonically descending towards the $K \to 0$ limit, which is in sharp contrast with the $G=2$ case. Reiterating the above argumentation for results presented in Fig.~\ref{square}, we thus confirm that larger group sizes can alter the effective interaction topology, in turn substantially affecting the evolution of cooperation in public goods games.

We note that our observations are not restricted to lattices but remain valid also for other interaction topologies having a low clustering coefficient. For example, we have employed a random regular graph with the degree of each player equal to $z=4$, whereby the lack of overlapping triangles and the directly related negligible clustering coefficient result in a similar noise-dependence of cooperation as depicted in Fig.~\ref{square}. Namely, using $G=2$ cooperators are optimally sustained at a finite value of $K$, yet this changes sharply if larger groups (\textit{e.g.} $G=5$) are considered. In the latter case, as we have observed for the above-mentioned lattices, the deterministic limit ($K \to 0$) yields an optimal environment for the evolution of cooperation at a given value of $r<G$. Importantly, since the reasoning of cooperation promotion via overlapping triangles for the prisoner's dilemma game (see \textit{e.g.} Fig. 33 in \cite{szabo_pr07}) does not apply for the public goods game, this work thus extends the generality of observations related to the graph type dependent phase transitions in the low and high $K$ limits within evolutionary games.

\begin{figure}
\centerline{\epsfig{file=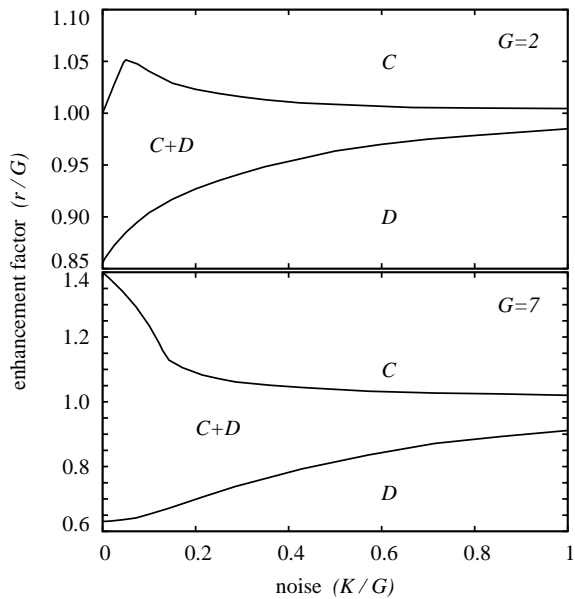,width=8.5cm}}
\caption{$K-r$ phase diagrams for the triangular lattice at $G=2$ (top) and $G=7$ (bottom). Note that the approach of the lower phase boundary towards the $K \to 0$ limit is independent on the group size $G$.}
\label{hexag}
\end{figure}

It is interesting to reverse our argumentation and test the impact of different value of $G$ when the interaction graph contains overlapping triangles. We show in Fig.~\ref{hexag} the full $K-r$ phase diagram for the triangular lattice ($z=6$) at $G=2$ (top) and $G=7$ (bottom). Note that this two-dimensional lattice is made up of overlapping triangles, and hence the transitions reported above for increasing group size on graphs with zero or negligible clustering coefficient should not be observable in this case. In particular, since overlapping triangles are inherent to the triangular lattice, already pairwise interactions ($G=2$) preclude the existence of an optimal noise intensity. The outlay of the upper phase boundary in the top panel of Fig.~\ref{hexag} fully confirms this expectation, which is also in agreement with results reported earlier for the prisoner's dilemma game \cite{szabo_pre05}. Accordingly, when increasing the group size to $G=7$ the qualitative features of the $C+D \rightarrow D$ transition line are preserved, and only the critical values of $r$ shift. For both $G=2$ and $G=7$ the lower phase boundary is monotonically descending towards the $K \to 0$ limit, which confirms the fact that the change in the effective interaction topology brought about by increasing group size can be observed only on graphs having a low clustering coefficient, and that only in the latter case, and if governed by pairwise interactions, the group size determines the noise dependence of the evolution of cooperation in spatial public goods games.

\section{Summary}

Summarizing, we have studied the evolution of cooperation in public goods games with different group sizes on various regular graphs. Starting deliberately with lattices that lack overlapping triangles, we have shown that pairwise interactions result in qualitatively similar behavior as was reported earlier for the prisoner's dilemma game \cite{hauert_c03, szabo_pre05}. In particular, in this case there exists an intermediate uncertainty governing the process of strategy adoptions $K$ by which the lowest enhancement factor $r$ still warrants the survival of at least some cooperators. By increasing the group size, we have demonstrated that the features characteristic for pairwise interactions vanish and become qualitatively identical to what was observed previously on lattices that do incorporate overlapping triangles, such as the kagome lattice. Since in fact the actual interaction topology remains unaffected by the different group sizes, we have argued that the differences in the evolution of cooperation are due to an effective transition of the interaction topology, which is brought about by joint memberships of players within the larger groups, resulting in indirect linkages among them. Identical results can also be obtained for increasing group sizes on random regular graphs, which essentially also have a low clustering coefficient and thus largely lack overlapping triangles. The latter results confirm our conjecture that the noise dependence of cooperation becomes qualitatively similar in public goods games if the group size is sufficiently large for a wide class of regular connectivity structures.

\begin{acknowledgments}
The authors acknowledge support from the Hungarian National Research Fund (grant K-73449), the Slovenian Research Agency (grant Z1-2032-2547), the Bolyai Research Grant, and the Slovene-Hungarian bilateral incentive (grant BI-HU/09-10-001).
\end{acknowledgments}


\end{document}